# Electric quadrupole interaction in cubic BCC α-Fe

A. Błachowski[1], K. Komędera[1], K. Ruebenbauer[1*], G. Cios[2], J. Żukrowski[2], and R. Górnicki[3]

[1]Mössbauer Spectroscopy Division, Institute of Physics, Pedagogical University
*ul. Podchorążych 2, PL-30-084 Kraków, Poland*

[2]AGH University of Science and Technology,
Academic Center for Materials and Nanotechnology
*Av. A. Mickiewicza 30, PL-30-059 Kraków, Poland*

[3]RENON
*ul. Gliniana 15/15, PL-30-732 Kraków, Poland*

[*]Corresponding author: sfrueben@cyf-kr.edu.pl

**Abstract**

Mössbauer transmission spectra for the 14.41-keV resonant line in $^{57}$Fe have been collected at room temperature by using $^{57}$Co(Rh) commercial source and α-Fe strain-free single crystal as an absorber. The absorber was magnetized to saturation in the absorber plane perpendicular to the γ-ray beam axis applying small external magnetic field. Spectra were collected for various orientations of the magnetizing field, the latter lying close to the [110] crystal plane. A positive electric quadrupole coupling constant was found practically independent on the field orientation. One obtains the following value $V_{zz} = +1.61(4) \times 10^{19}$ Vm$^{-2}$ for the (average) principal component of the electric field gradient (EFG) tensor under assumption that the EFG tensor is axially symmetric and the principal axis is aligned with the magnetic hyperfine field acting on the $^{57}$Fe nucleus. The nuclear spectroscopic electric quadrupole moment for the first excited state of the $^{57}$Fe nucleus was adopted as +0.17 b. Similar measurement was performed at room temperature using as-rolled polycrystalline α-Fe foil of high purity in the zero external field. Corresponding value for the principal component of the EFG was found as $V_{zz} = +1.92(4) \times 10^{19}$ Vm$^{-2}$. Hence, it seems that the origin of the EFG is primarily due to the local (atomic) electronic wave function distortion caused by the spin-orbit interaction between effective electronic spin **S** and incompletely quenched electronic angular momentum **L**. It seems as well that the lowest order term proportional to the product **L**•**λ**•**S** dominates, as no direction dependence of the EFG principal component is seen. The lowest order term is isotropic for a cubic symmetry as one has $\lambda = \lambda \mathbf{1}$ for cubic systems with the symbol **1** denoting unit operator and $\lambda$ being the coupling parameter.

## 1. Introduction

In principle one cannot expect any electric field gradient (EFG) on the iron sites (nuclei) in the α-Fe having BCC structure. However, one has ferromagnetic order with the iron magnetic moments tending to be aligned with one of the main axes of the chemical unit cell i.e. some magnetic anisotropy within the crystal (magnetic domain) with the easy axis of magnetization being one of the main axes of the chemical cell. On the other hand, a hard (difficult) axis is one of the cell diagonals. This weak anisotropy and corresponding magneto-elastic effects in a form of the magnetostriction are due to the incomplete quench of the orbital angular momentum as the gyromagnetic factor differs from the pure spin electronic factor by about 0.1. The resulting spin-orbit coupling may induce a perturbation of the (atomic) electronic wave function leading to the electric quadrupole interaction with the nuclear electric quadrupole moment via magnetically induced electric field gradient (EFG) tensor [1 - 4]. The EFG is likely to be the axially symmetric tensor with the principal axis aligned with the iron magnetic moment, and hence, with the hyperfine magnetic field on the iron nucleus. An attempt to calculate magnetic anisotropy from the first principles was only partly successful in the case of α-Fe due to the smallness of this relativistic effect [5]. Hence, there is little hope to get reliable results for corresponding EFG by calculations. The EFG has been observed previously in the α-Fe applying 14.41-keV Mössbauer transition in $^{57}$Fe [6, 7]. These early measurements were performed at room and liquid helium temperatures on the rolled polycrystalline foils. Some positive quadrupole coupling constants were found in the majority of cases. Later on some measurements were performed on a magnetized single crystal and some difference in the coupling constant for two measured directions <100> and <111> was found [8]. However, this finding appeared hard to reproduce and the role of the material strain was emphasized [9]. We have observed previously small positive coupling constant at room temperature as well [10] by using standard as-rolled iron calibration foils. Hutchison *et al.* [11] used modulated adiabatic passage on oriented $^{59}$Fe radioactive nuclei (MAPON) method at extremely low temperature. Measurements were performed by using single crystal and in two directions of the magnetization i.e. for <100> and <111> directions. No significant anisotropy of the coupling constant was found and the claim was made that the coupling constant corresponds to the positive principal component of the EFG being about +1.8 x $10^{19}$ Vm$^{-2}$. However, the claim about the sign of the principal component was later retracted [12]. The EFG was observed for diluted 5d-impurities in cubic hosts – in particular in the α-Fe [13, 14]. Calculations within spin polarized relativistic Korringa-Kohn-Rostocker (SPR-KKR) approach reproduce quite well experimental data for 5d- and 4d-impurities without resorting to the lattice relaxation [15]. Such result is a strong indication that the phenomenon is mainly caused by redistribution of the local electrons surrounding nucleus in question. Results are definitely less understood for 3d-impurities in 3d cubic hosts. For review see Ref. [11]. In particular the case of Fe in α-Fe seems unclear.

Hence, it seems interesting to have a closer look at this phenomenon in a more systematic way making measurements on the α-Fe single crystal with different orientations of magnetization.

## 2. Experimental

The single crystal of α-Fe with the natural isotopic composition was purchased from GoodFellow in the form of a disk having 10 mm diameter and 2 mm thickness. The crystal was obtained as strain-free and of the 5N nominal purity. The <110> axis was nearly perpendicular to the disk surface. A slice of about 0.3 mm thickness was cut parallel to the disk surface using Unipress WS-22 wire thaw equipped with 60 μm diameter tungsten wire.

Slurry made of mineral oil and SiC powder of 0.3 μm diameter was used as the abrasive medium. The slice was polished to the final average thickness of about 40 μm via mechanical polishing using SiC papers with gradation from 320 to 2000. The orientation of the crystal was performed by means of the Empyrean diffractometer by Panalytical set to Bragg-Brentano geometry and applying Co-Kα radiation (working at 40 kV and 40 mA) with *chi* and *phi* steps of the stage set to $1^o$. Prior to orientation the crystal was glued to the lucite support of 1 mm thickness and later on covered with nearly identical lucite disk. The active diameter was set to 6.5 mm in the center of the sample by means of the brass collimator. The whole ensemble was magnetized to saturation applying field of about 0.18 T in the sample plane with a provision to rotate the sample around the γ-ray beam axis, the latter being perpendicular to the sample surface. Two stacks of permanent Nd-Fe-B magnets coupled to the soft steel yokes were used to provide magnetizing field. Details of the geometry are shown in Figure 1. For geometry applied one can scan sample almost within [110] plane going nearly through the easy and hard axes. The γ-ray beam axis was oriented horizontally.

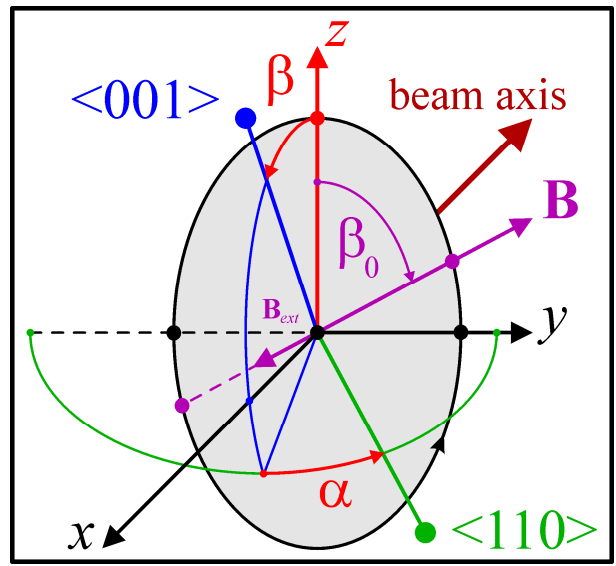

**Fig. 1** Details of the experimental geometry are sketched. Note that the sample could be always oriented in such way that one has $\beta \geq 0$ as shown. For $\beta = 0$ one can always define right-handed system {*xyz*} with $\alpha \geq 0$. Angles shown ($\beta, \alpha, \beta_0$) are positive. Arrow on the edge of the image of the sample central plane marks rotation of the sample versus applied field $\mathbf{B}_{ext}$. The symbol **B** denotes the hyperfine field on the $^{57}$Fe nucleus. The following values for the respective angles were found during crystal orientation $\alpha = -0.85^o$ and $\beta = +1.00^o$.

A fresh commercial $^{57}$Co(Rh) source obtained from Ritverc G.m.b.H. was used. The source had 6 μm thickness and active diameter of 8 mm. The nominal activity amounted to 50 mCi. The source was covered with a thin high purity Be window and the source holder was made of titanium. The source stayed in a double wall collimator with the internal collimator made of ARMCO and external from lead. The collimator was equipped with a window made of thin organic adhesive tape. The collimator with the attached magnetic circuit containing iron sample was rigidly fastened to the transducer body. Source and absorber remained at ambient temperature (about 24 $^o$C) being nearly the same for the source and absorber. Narrow beam geometry was applied with the detector being far away from the source/absorber assembly. A Kr-filled proportional detector obtained from LND Inc. was used in the linear regime as far as the count-rate and pulse amplitudes are considered. A detector was shielded by the double layer shield with the internal layer made of ARMCO and external of lead and it was equipped with the high purity Be window. The RENON MsAa-4 Mössbauer spectrometer was used to collect Mössbauer spectra in the 14.41-keV photo-peak and escape peak of the detector simultaneously. A velocity scale was calibrated by means of the Michelson-Morley interferometer equipped with a single mode metrological quality He-Ne laser (red light) rigidly attached to the back of the transducer. A single longitudinal mode operation was assured applying small magnetic field oriented perpendicular to the axis of the discharge tube by means of the permanent magnet. The JDS Uniphase Corp. provided He-Ne laser. The moving corner prism was attached just to the back of the source extender made of carbon fiber, while the random noise was applied to the reference prism using small piezoelectric transducer to avoid phase locking. The light beams were separated spatially to avoid multiple interference patterns and aligned with the transducer axis. Corner prisms and photo-detector were tilted to avoid formation of cavities leading to the multiple scattering of light. The beam expander focused "to infinity" was applied to assure better mechanical stability and beam parallelism. Wavelength of the laser monochromatic light was corrected for the air pressure and temperature in the vicinity of the moving prism. A triangular reference velocity function with round-corners was used and the spectrometer repetition frequency was about 5 Hz. The average position of the moving source was stabilized.

In order to check linearity of the Mössbauer transducer the pick-up coil signal was collected for many periods of the spectrometer and the average was fitted to the second order polynomial upon being folded like the Mössbauer spectrum. The pick-up coil sensitivity was 80.4 mVmm$^{-1}$s. Figure 2 shows results of this test performed simultaneously with the Mössbauer measurements. One can conclude that relative deviations of the velocity scale from the reference velocity are smaller than 0.003 % within full useable velocity range.

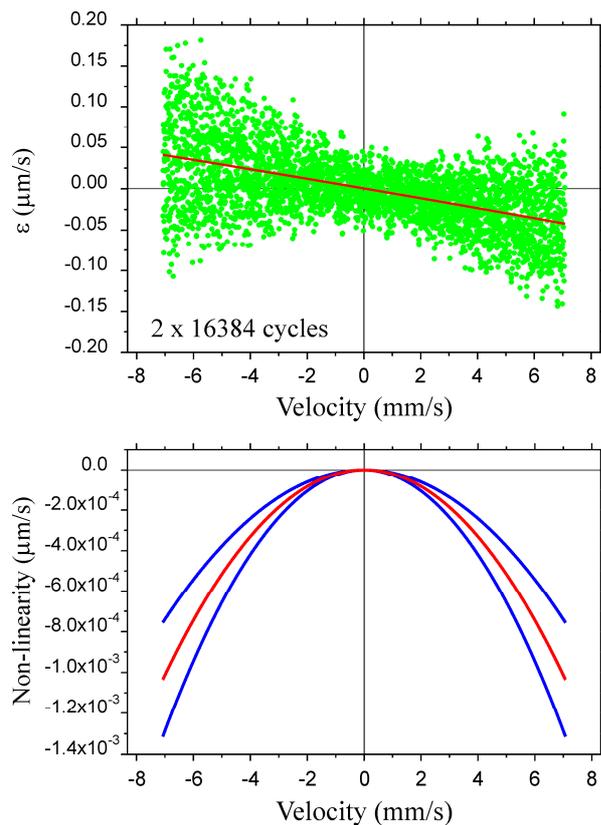

**Fig. 2** Linearity test of the transducer performed for two series of cycles composed of 16384 cycles each (upper part). The symbol $\varepsilon$ denotes deviation from linear reference function. The lower part shows non-linear terms with respective error bands i.e. deviation of the solid (red) line (shown in the upper part) from the straight line, as non-linear terms could be responsible for deviations of the line positions from the "true" positions. Experimental points cover about 85 % of the velocity range and there are 3482 points shown.

Data were summed over detector windows and folded according to the calibration data. They were subsequently fitted within transmission integral approach. The Mosgraf-2009 suite was used to process data [16].

Additionally, the spectrum of the 4N purity as-rolled iron foil of natural isotopic composition was collected with high statistics at nearly the same geometry, albeit without external field. The foil had thickness of 25 μm with the total diameter of 16 mm and was obtained from GoodFellow. The foil was sandwiched between two mylar windows having 0.1 mm thickness each. Both windows were aluminized on both sides each using high purity aluminum. Single crystal spectra were accumulated for approximately 24 h each, while the spectrum of the as-rolled foil was taken for almost 4 days after collecting single crystal spectra. All measurements constituted single series of uninterrupted measurements lasting 45 days. Spectra were collected for 4096 channels each and upon folding 2047 channels remained. Calibration runs were performed simultaneously with the Mössbauer measurements and collection of the γ-ray spectra with appropriate detector windows included (both data sets were collected in 4096 channels each). All data channels were of 32-bit capacity each.

### 3. Data evaluation and results

Data were fitted to the standard transmission integral expression. For each spectrum the following parameters were fitted. The number of counts per data channel far-off the resonance (baseline), an effective recoilless fraction of the source with the background in the detector windows taken into account, the source line width being equated to the absorber line width, and dimensionless resonant absorber thickness were variables. Additionally, the magnetic field acting on the $^{57}$Fe nucleus, total spectral shift, the parameter $g_{11}^{111}$, and the (quadrupole) coupling constant $A_Q$ were varied. The field was calculated using the following scale $\mu_N g_g (c/E_0) = +0.1187843$ mm s$^{-1}$T$^{-1}$ for the ground nuclear state having spin and parity $I_g^{(\pi)} = \frac{1}{2}^{(-)}$. The symbol $\mu_N$ denotes nuclear magneton, $g_g$ stands for the nuclear gyromagnetic factor in the ground nuclear state, $c$ denotes speed of light in vacuum, and $E_0$ is the transition energy from the ground to the first excited nuclear state in $^{57}$Fe (14.410 keV). The ratio of gyromagnetic factors was set to $g_e / g_g = -0.5714$, where the symbol $g_e$ stands for the gyromagnetic factor in the first excited nuclear state of the spin and parity $I_e^{(\pi)} = \frac{3}{2}^{(-)}$. A transition is practically of the pure M1 character. The parameter $g_{11}^{111}$ is responsible for the corrections to the relative line intensities in the resonant absorption cross-section, the latter being due to the non-random distribution of the magnetic field directions with respect to the γ-ray beam [17]. For the field being perpendicular to the beam axis one obtains in the present context $g_{11}^{111} \rightarrow \frac{1}{2}$, for random case one has $g_{11}^{111} = 1$, while for the field being aligned with the beam this parameter diverges $g_{11}^{111} \rightarrow +\infty$. The quadrupole coupling constant takes on the form $A_Q = \frac{eQ_e (c/E_0) V_{zz}}{4I_e (2I_e - 1)}$ with the symbol $e > 0$ denoting positive elementary charge, and the symbol $I_e = \frac{3}{2}$ standing for the spin of the excited nuclear state. The symbol $Q_e$ stands for the spectroscopic nuclear electric quadrupole moment in the first excited state of $^{57}$Fe. It takes on the following value for this particular nuclear state $Q_e = +0.17$ b [18]. Finally, the symbol $V_{zz}$ stands for the principal component of the EFG tensor (diagonal and axially symmetric here with the principal component aligned with the magnetic field acting on the nucleus). For the high statistics spectrum of the polycrystalline iron foil ratio of the gyromagnetic factors $g_e / g_g$ was made variable, the source and absorber line widths were fitted separately, and the second and fourth order polynomial corrections to the baseline were introduced and fitted separately each. These corrections follow from the fact, that source moves in a correlated periodic fashion along the γ-ray beam axis. One has to note, that for magnetically saturated sample the absolute value of the field acting on the nucleus equals $B - B_{ext}$ with $B = |\mathbf{B}|$ and $B_{ext} = |\mathbf{B}_{ext}|$ under obvious here assumption that the following relationship holds $B_{ext} << B$.

Left panel of Figure 3 shows typical spectrum of the single crystal with the differential plot obtained by subtraction of the fitted function from the experimental data and dividing resulting difference by the statistical error of the experimental data. A small and broad (unaccounted for) dip in the center of the spectrum is due to the spurious iron dissolved in the Be window of the detector. This feature has no influence on the essential results. Right panel shows spectrum of the as-rolled iron foil with the corresponding differential plot. Some additional (unaccounted for) small deviations are due to the tiny variation of the magnetic hyperfine fields in the Bloch walls between magnetic domains. The foil is only partially magnetized in the foil plane by the shape anisotropy, while the crystal is completely magnetized by the applied field and the part of the crystal exposed to the γ-ray beam is a single magnetic domain. Bloch walls are pinned by the grain boundaries in the polycrystalline foil with slightly different isomer shift. Hence, there are some correlated distributions of the isomer shift and hyperfine field in the foil. These additional features are insignificant as far as important parameters are concerned. The second (lower) differential plot is obtained by fitting the foil spectrum in the same way as previously, albeit with the quadrupole coupling constant fixed at zero value. It is obvious that the non-zero quadrupole coupling constant is necessary to fit properly data.

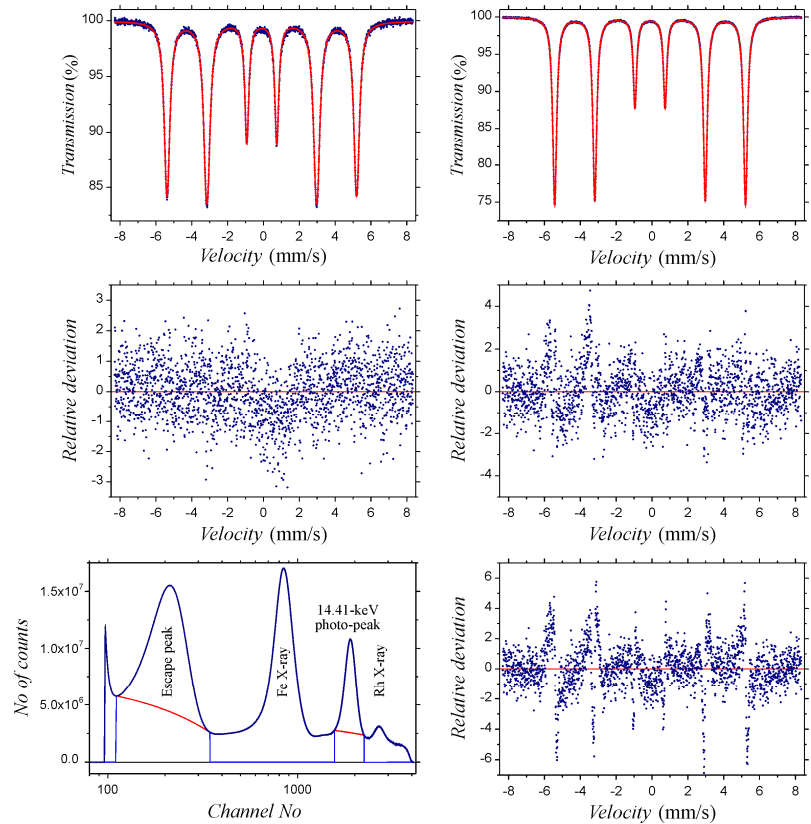

**Fig. 3** Left panel shows the spectrum of the magnetized single crystal obtained for the angle $\beta_0 = +10^\circ$ during run with the increasing angle $\beta_0$ and corresponding differential plot described within text. The solid curve represents fitted function. Right panel shows the spectrum of the as-rolled foil obtained without external field and fitted to the complete set of parameters (solid curve shows fitted function). The upper differential plot corresponds to above fit, while the lower is shown for fit with the same set of variables (parameters) except the quadrupole coupling constant, the latter set to zero. Spectra are normalized by respective baselines. The lowest left panel shows γ-ray spectrum (acquired in 4096 data channels) of the foil with detector windows used to collect Mössbauer spectra and estimated divisions between "signal" and "background" by means of simple linear interpolation.

Line broadening for single crystal spectra in comparison with the foil spectrum is due to the much greater resonant thickness of the crystal in comparison with the foil. On the other hand, lower effect for crystal in comparison with the foil is primarily caused by the enhanced multiple Compton scattering of the 121.91-keV line in a thicker layer of rather heavy iron leading to the increased background.

Altogether forty-one Mössbauer spectra were obtained for the single crystal magnetized to saturation with the angle $\beta_0$ varying from zero to the positive right angle, later on decreasing to the negative right angle, and again increasing to the value of zero. Essential parameters are shown in Figure 4. The field acting on the $^{57}$Fe nucleus is perturbed by the magnetizing field and therefore is not shown. The magnetizing field slightly varies within the bulk of the sample (being however a constant vector within sample) due to the imperfections of the sample edge. Applied collimator shields above imperfections. Basically the same statement applies to the as-rolled foil as the field acting on the $^{57}$Fe nucleus is slightly perturbed within Bloch walls and some of them are exposed to the transmitted γ-ray beam.

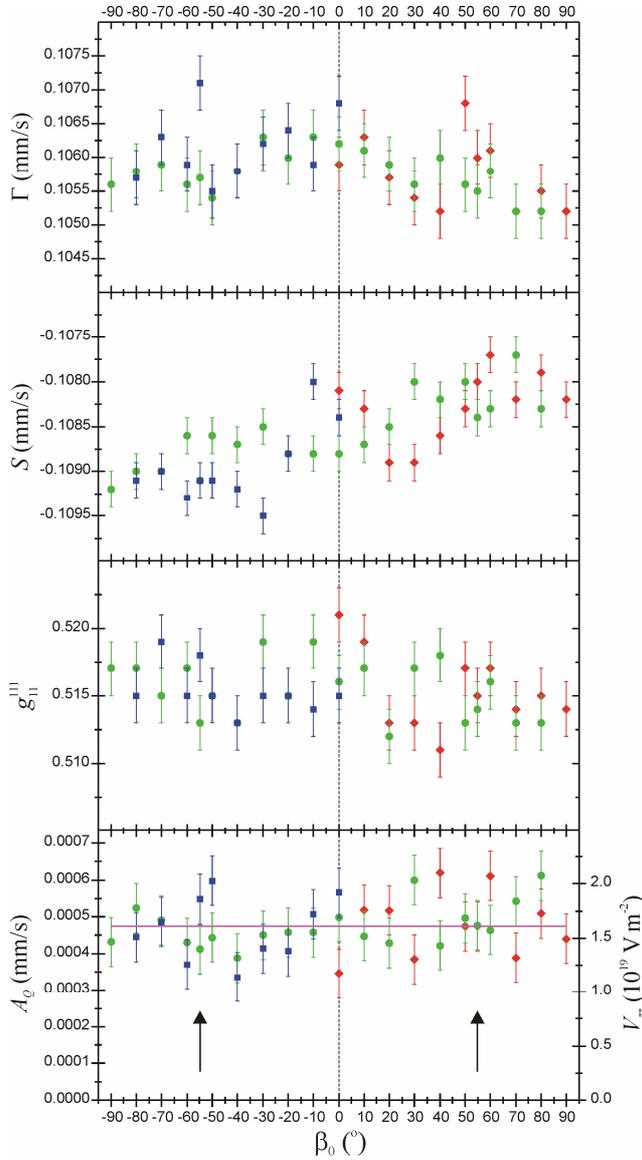

**Fig. 4** Essential parameters derived from single crystal spectra. Diamonds correspond to the increasing angle $\beta_0$ from zero to the positive right angle. Circles correspond to the subsequent decreasing angle $\beta_0$ from the positive right angle to the negative right angle, and finally squares correspond to the final runs with the angle $\beta_0$ increasing from the negative right angle to the zero value. The symbol $\Gamma$ stands for the line width (line widths of the absorber and source were equated one to another), the symbol $S$ stands for the total spectral shift, the symbol $g_{11}^{111}$ stands for the relative line intensity correction due to the non-random field orientation, and the symbol $A_Q$ denotes quadrupole coupling constant. The right hand scale of this last (lowest) panel shows corresponding principal component of EFG $V_{zz}$. The solid horizontal line of this panel shows weighed average over all measurements. Arrows show directions closest to the directions equivalent to the direction <111> i.e. for $\beta_0 = \pm 54.7°$. The maximum possible misalignment of the crystal does not exceed $3°$ on the unit sphere.

The weighed average for the principal component of the EFG in single crystal amounts to $V_{zz} = +1.61(4) \times 10^{19}$ Vm$^{-2}$ with the corresponding weighed average coupling constant being $\langle A_Q \rangle = +0.47(1)$ μm s$^{-1}$. One obtains the following values of the essential parameters for the as rolled foil $\Gamma_S = 0.106(2)$ mm s$^{-1}$ (source line width), $\Gamma_A = 0.106(2)$ mm s$^{-1}$ (absorber line width), $S = -0.10906(4)$ mm s$^{-1}$ (total spectral shift), $g_{11}^{111} = 0.7027(5)$ (line intensity correction), and $V_{zz} = +1.92(4) \times 10^{19}$ Vm$^{-2}$ (principal component of EFG). Corresponding quadrupole coupling constant amounts to $A_Q = +0.56(1)$ μm s$^{-1}$. Hence, the outer lines of the spectrum are shifted versus remaining lines by $+3.36(6)$ μm s$^{-1}$. The shift is about $3 \times 10^{-4}$ of the splitting of the outer lines (about 0.39 of the adjacent data channels distance). The value of the magnetic field seen on the $^{57}$Fe nucleus in the as-rolled foil amounts to 33.034(1) T with $g_e / g_g = -0.57100(3)$. One can see that the magnetizing field practically saturated single crystal for all orientations. A difference in the quadrupole coupling constants of the single crystal and as-rolled foil seems barely significant taking into account smallness of the observed effect. Slightly higher principal component of the EFG in a foil as compared to the single crystal (by $+0.31(6) \times 10^{19}$ Vm$^{-2}$) indicates that strain induced EFG has positive principal component as well. A total spectral shift (practically isomer shift here) is typical for currently manufactured $^{57}$Co(Rh) sources based on the high purity Rh matrix. It means that the electron density on $^{57}$Fe nucleus is lower by $0.375$ el. (a.u.)$^{-3}$ in Rh foil in comparison with α-Fe (rolled foil) in the vicinity of room temperature [18]. Parameters displayed in Figure 4 do not have significant correlation with the angle $\beta_0$ except total (isomer) shift $S$. The last (almost linear) correlation is likely to be due to the non-radial inhomogeneity of the specific activity in the source (in the plane perpendicular to the beam axis and measured relative to the beam axis) combined with similar inhomogeneity of the single crystal absorber resonant thickness.

## 4. Conclusions

It appears that ferromagnetic order in the cubic BCC α-Fe breaks cubic electron charge symmetry around iron nucleus. Symmetry is broken via the spin-orbit coupling between effective electron spin **S** and incompletely quenched effective electron angular orbital momentum **L**. The coupling **L**•**λ**•**S** seems direction independent i.e. the principal component of EFG does not depend upon orientation of the electronic magnetic moment. Hence, the coupling tensor takes on the form $\boldsymbol{\lambda} = \lambda \mathbf{1}$ with the symbol **1** denoting unit operator and $\lambda$ being the coupling parameter. Such form of the coupling tensor is consistent with the cubic symmetry of the crystal. On the other hand, the spin-orbit interaction leads to the axial distortion of the electronic charge around nucleus with the symmetry axis being direction of the electronic magnetic moment – the hyperfine magnetic field on the iron nucleus. A distortion creates small axially symmetric EFG on the iron nucleus with the principal axis aligned with the hyperfine field. The principal component of the EFG is positive. Hence, one can conclude that electronic states with the highest magnetic quantum numbers (in the absolute sense) are more occupied (have more significant amplitudes) in comparison with remaining states due to the spin-orbit interaction for a quantization axis chosen along the hyperfine field.

It seems that extended defects introduced by rolling have minor effect, as the quadrupole coupling constants are similar in the as-rolled polycrystalline foil and in the strain-free single crystal. Strain induced residual EFG has principal component of the same positive sign as the EFG due to the spin-orbit coupling. Hence, it is likely that effective spin-orbit coupling constant $\lambda$ is enhanced by the residual strain.

The principal component of EFG obtained by us at room temperature is not much different from the value obtained practically in the ground state of the system by Hutchison *et al.* [11]. Hence, a temperature dependence of the EFG is rather weak and probably follows overall internal sample magnetization. Some measurements at higher temperature are welcome to solve this problem – particularly closer to the rather high Curie temperature of the α-Fe established as 1043 K.

The Mössbauer spectroscopy of the 14.41-keV line in $^{57}$Fe is capable to obtain similar accuracy as MAPON method provided the quadrupole interaction (axial EFG with the principal component aligned with the hyperfine field) is a small perturbation to the dominant magnetic dipole interaction free of hyperfine anomaly. The Mössbauer spectroscopy has the advantage of being able to measure sign of the coupling constant (sign of the principal EFG component), it is easy to obtain multiple spectra for various orientations of the magnetization in a single crystal, and measurements could be in principle performed for wide temperature range. One does not need to introduce radioactive nuclei into the sample.

**Acknowledgments**

Dr. Sc. Michael I. Oshtrakh, Ural State Technical University, Ekaterinburg, Russian Federation is warmly thanked for inspiring this research. This work was partly supported by the National Science Center of Poland, Grant No DEC-2011/03/B/ST3/00446 and by the European Regional Development Fund under the Infrastructure and Environment Programme. Development of the MsAa-4 spectrometer was partly financed by the Polish Ministry of Science and Higher Education under the Grant No R15-002-03.